\documentclass[conference]{IEEEtran}
\IEEEoverridecommandlockouts
\usepackage{cite}
\usepackage{amsmath,amssymb,amsfonts}
\usepackage{algorithmic}
\usepackage{graphicx}
\usepackage{textcomp}
\usepackage{xcolor}
\def\BibTeX{{\rm B\kern-.05em{\sc i\kern-.025em b}\kern-.08em
    T\kern-.1667em\lower.7ex\hbox{E}\kern-.125emX}}

\usepackage[]{algorithm2e}
\SetAlFnt{\small\sffamily}

\usepackage{booktabs} 
\newcommand{\dorien}[1]{\textcolor{black}{#1}} 
 
\usepackage[numbers]{natbib}
\usepackage{graphicx}
\newcommand{\Csharp}{%
  {\settoheight{\dimen0}{C}C\kern-.05em \resizebox{!}{\dimen0}{\raisebox{\depth}{\#}}}}

\title{A novel music-based game with motion capture to support cognitive and motor function in the elderly\\
\thanks{We thank the 
students supported under the UROP project ``Developing A Movement-Based Music Game For Preventive Healthcare And Well-Being In The Elderly''. They include Xuexuan Zhou who led the Unity implementation, and Kenny Soh Chi Tong and Tan Li Yang who helped with gesture implementation.
This work is supported in part by SUTD under Grant
No.:SUTD SRG ISTD 2017 129 and SUTD Game Lab.}
}

\author{\IEEEauthorblockN{Kat Agres\IEEEauthorrefmark{1}\IEEEauthorrefmark{2}, Simon Lui\IEEEauthorrefmark{3}, \& Dorien Herremans\IEEEauthorrefmark{2}\IEEEauthorrefmark{3}}

\IEEEauthorblockA{\IEEEauthorrefmark{1}\textit{Yong Siew Toh Conservatory of Music}, 
\textit{National University Singapore}, 
Singapore}

\IEEEauthorblockA{\IEEEauthorrefmark{2}\textit{Social \& Cognitive Computing Department}, \textit{Institute of High Performance Computing, A*STAR}, 
Singapore}

\IEEEauthorblockA{\IEEEauthorrefmark{3}\textit{Information Systems, Technology \& Design}, \textit{Singapore University of Technology \& Design}, 
Singapore \\
muskra@nus.edu.sg, nomislui@gmail.com, dorien\_herremans@sutd.edu.sg}
}
\usepackage{url}

\begin{document}
\IEEEpubid{\begin{minipage}{\textwidth}\ \\[12pt]
Preprint accepted to the IEEE Conference on Games 2019, London, UK. 
\end{minipage}}
\maketitle

\sloppy




  

\begin{abstract}
This paper presents a novel game prototype that uses music and motion detection as preventive medicine for the elderly.
Given the aging populations around the globe, and the limited resources and staff able to care for these populations, eHealth solutions are becoming increasingly important, if not crucial, additions to modern healthcare and preventive medicine. Furthermore, because compliance rates for performing physical exercises are often quite low in the elderly, systems able to motivate and engage this population are a necessity. Our prototype uses music not only to engage listeners, but also to leverage the efficacy of music to improve mental and physical wellness. The game is based on a memory task to stimulate cognitive function, and requires users to perform physical gestures to mimic the playing of different musical instruments. To this end, the Microsoft Kinect sensor is used together with a newly developed gesture detection module in order to process users' gestures. The resulting prototype system supports both cognitive functioning and physical strengthening in the elderly. 
\end{abstract} 

\begin{IEEEkeywords}
music, serious games, motion detection, motor skills, cognitive function, eHealth, telehealth, aging population
\end{IEEEkeywords}




\sloppy

\section{Introduction}

Many countries are facing the looming crisis of an aging population. In North America and Europe, for example, more than one in five people were aged 60 years or over in 2017, according to a recent report by the United Nations~\cite{un2017}. 
Globally, the number of people aged 80 or over is expected to increase more than three times by 2050, rising from 137 million in 2017 to 425 million~\cite{un2017}. 
This number far surpasses the capacity of hospitals, clinics, and care centers (e.g., \cite{chan2011singapore}), and carries heavy implications for the incidence of age-related cognitive and motor impairments.
In addition, many countries face a decreasing old-age support ratio\footnote{This ratio equals the number of residents aged 20-64 years to support each resident aged 65 and over, and is down in Singapore, for example, from 13.5 in 1970 to only 5.1 in 2018, according to the Dept. of Statistics Singapore.}, which will place an increasingly large burden on already-stressed health care systems across the globe. 

This wide-spread concern begs the questions: 
\textit{What strategic measures can be taken in preparation? How can technology play a central role in preventive medicine? And how can we support the elderly in maintaining their motor and cognitive functions?} Preventive medicine, including music-based eHealth technologies, may play a crucial role in the answer.

Music therapy has been shown to be effective in supporting both cognition and mental health. For example, music interventions have repeatedly been shown to improve a variety of cognitive functions, including executive control~\citep{thaut2009neurologic}, attention~\citep{kim2008effects}, verbal recall \citep{gfeller1983musical, wolfe1993use}, working and autobiographical memory \citep{prickett1991use, simmons2010music, irish2006investigating}. Research examining the relationship between musical activity and cognitive aging has suggested that practicing music can lead to the transfer of music-related cognitive abilities to non-musical cognitive functions, improving nonverbal memory and executive processing \citep{hanna2011relation}. In addition, musical interventions have been shown to significantly decrease anxiety and depression in the elderly \citep{thaut2009neurologic,irish2006investigating, khalfa2003effects,white1992music}. Given the many benefits of music-based interventions, we aim to explore the efficacy of music technology to help prevent motor and cognitive decline in an elderly population.

One current challenge is that, although traditional music therapy interventions are effective, they are also costly, because a certified therapist must be employed (and the therapist often only provides 1-to-1 care or therapy for small groups of patients). \textit{Music-based eHealth technology has the ability to remedy these shortcomings by cutting costs and allowing widespread usage}. 
Given the forthcoming ``silver tsunami'', there is not only a need for preventive healthcare, but increasingly for \textit{remote} preventive care, or tele-preventive interventions. We address the urgent need for \textit{engaging} preventive medicine by presenting here a prototype music system to improve cognitive and physical health and wellness for the elderly. The game incorporates music for its health-related aspects, but also for its ability to help motivate and engage users, as the elderly often struggle with compliance~\cite{brand1974medical}, and have to overcome more barriers to exercise than their younger counterparts~\cite{guest2002promoting}.  
The serious game developed in this research 
could provide a long-term health solution by empowering participants to take preventive medicine into their own hands.


\section{Existing systems For Preventive Medicine} 

There has been a steady rise in the number of computer games for health care (i.e., serious games) since 2008~\cite{kharrazi2012scoping}. While there currently exist a number of games for healthcare applications, few games have been developed for older adults. Indeed, \citet{kharrazi2012scoping} has suggested widening the demographics for serious games to include the elderly, which is the target population of the prototype described here. 

The use of games for improving cognition in the elderly has been explored by a number of researchers. The work of \citet{ballesteros2014brain} examines the effect of using non-action games from the commercially available `Brain games' package by Luminosity\footnote{\url{http://lumosity.com}} with elderly participants. The results of their study indicate an improvement in some cognitive functions (e.g. processing speed, attention, immediate and delayed visual recognition memory), but not others (e.g. visuospatial working memory). The potential of real-time strategy video games with elderly is explored by \citet{basak2008can}, who observed improvements in task switching, reasoning, working memory, visual short term memory, and mental rotation. 
In a meta-review, \citet{toril2014video} compare the results of 20 different experimental studies executed between 1986 and 2013 and found that the use of video games for elderly has a positive effect on multiple cognitive functions such as memory, reaction time (RT), attention, and global cognition. \looseness=-1

While the systems mentioned above focus on an important issue regarding aging, cognitive decline, the trend towards increased longevity also creates concerns relating to physical decline. This is another area where games have been proven useful, often by integrating motion sensors such as the Microsoft Kinect\footnote{https://developer.microsoft.com/en-us/windows/kinect}. For example, games have been developed in the context of physical exercises for motor rehabilitation in stroke patients~\cite{chang2013kinect,chang2011kinect,beveridgerhythmic,agres2017icot}; balance rehabilitation~\cite{lange2011development}, and improving balance in the elderly~\cite{lai2013effects}. \looseness=-1

Our system includes a physical activity component during gameplay: performing movements to mimic the gestures of playing different instruments (see Section 3 for details). Gestures are detected using the Kinect sensor, which includes a color camera and a depth sensor. The Kinect device has proven to be a useful tool in clinical settings for movement evaluation~\cite{clark2012validity}, and can even parallel the functionality of a full multi-camera 3D motion system. 

The integration of a physical component into our game is well-aligned with research by \citet{laurin2001physical}, for example, who found a connection between increased physical activity and lower rates of cognitive impairment and dementia in the elderly. 
A study by \citet{kayama2014effect} integrated both cognitive and physical exercises in a game by letting the players solve a sudoku puzzle using entire body movements (similar to  Tai Chi Chuan) as a game controller. While the system was shown to be effective for improving cognitive function, the results with respect to physical improvements were mostly inconclusive. Similarly, \citet{anderson2012exergaming} found that using a static home training bicycle together with a virtual environment simulator prevents cognitive decline more effectively than cycling without this virtual environment. \looseness=-1

\begin{figure}[h] \centering
\includegraphics[width=.48\textwidth]{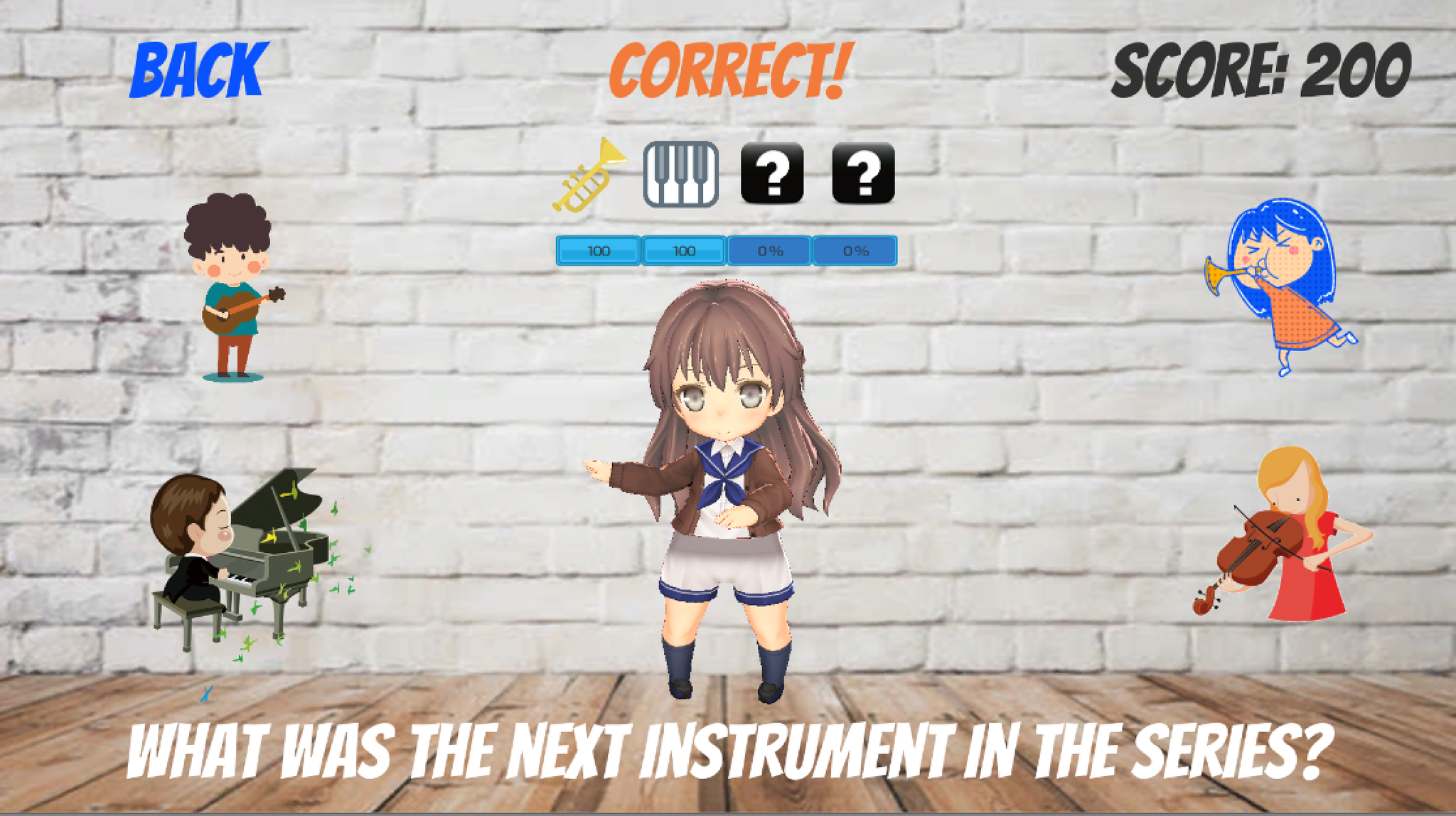}
\vspace{-.2cm}
\caption{Screenshot of the prototype system}
 \vspace{-.5cm}\label{fig:screenshot}
\end{figure}

The game system proposed in this research improves upon existing systems by creating a dedicated system for cognitive and motor support in the elderly that integrates the therapeutic aspects of music. Music is a complex domain that activates many cognitive resources and has shown to have a positive influence on mental health (e.g., by reducing depression and anxiety \cite{clair2008therapeutic}). In addition, the field of neurological music therapy (NRT) has drawn upon aspects of music perception, such as auditory motor synchronization (in which listeners' movements synchronize with an external rhythm), to create music-based therapies for motor control~\cite{thaut2005rhythm}. \looseness=-1 

Currently, only a handful of eHealth systems involve music in a meaningful manner. Of those that do exist, most are not tailored for preventative medicine (including the use of music as a motivational tool), and do not include gamification or actively incorporate music into gameplay. One of the few existing music-based eHealth tools is an app developed by \citet{ellis2015validated} that uses music to enhance movement in Parkinson's patients (however, this is not for preventive medicine). To the authors' knowledge, there are no dedicated systems that use music to both motivate physical movement \textit{and} stimulate cognitive function in the elderly. 
Our proposed prototype system utilizes music in a game to support cognitive and physical exercise in the elderly. The components of the system are described in detail in the next section. \looseness=-1





\section{Prototype system description}
The work described herein has several objectives. First, we aim for the system to be usable outside of clinical contexts (i.e., at home), and without the help of healthcare professionals. The game is centered around music, for its therapeutic benefits, as well as its power to motivate and engage listeners. The game itself requires users to make a series of gestures as part of a gamified memory task that is developed specifically to support motor and cognitive function in the elderly. \looseness=-1

\subsection{System overview}

We use the Kinect sensor with the Microsoft SDK 
for Kinect, which gives us the ability to perform real-time, full-body 3D motion capture with skeleton tracking and facial recognition~\cite{zhang2012microsoft}. The game itself was developed in Unity using \Csharp. Our system contains several components, including a tutorial, gameplay, and gesture detection; an overview of the system's architecture is available online\footnote{\url{http://katagres.com/musicgesturesgame}}. \looseness=-1

In the proposed game, the user first completes a tutorial to become familiar with the gestures required to play the game. The tutorial walks the user through a cartoon example of four gestures (one at a time), each associated with one of the four instruments used in the game: guitar, trumpet, keyboard, and violin. These gestures were selected because they require distinctively different types of movements during gameplay, and because several of the gestures require dissociative movements (i.e., exercises in which one limb performs a movement without the other limb concurrently performing the same or similar movement). These kinds of coordination movements can benefit cognitive functioning in the elderly~\cite{kwok2011effectiveness}.


Once the player is familiar with the gestures, the tutorial will randomly select one gesture for the user to perform. This process repeats until the user is able to produce the correct gesture at least 80\% of the time (accuracy is calculated based on a sliding window such that at least 4 out of the last 5 attempted gestures are correct).
In the future, this component of the system will be replaced with a machine learning (ML) gesture-recognition model: 
instead of ensuring that the user practices the gestures until 80\% accuracy is obtained, our future prototype will aim to learn and accurately classify the user's gestures for each instrument type. This approach will also provide a degree of customizability for elders who need to modify the gestures slightly due to physical constraints. 

Once the user has successfully completed the tutorial, the actual game begins. On each trial of the game, the user hears a melody comprised of a series of four excerpts, where each excerpt is played by a different solo instrument. For example, the trumpet will play, followed by the piano, the guitar, and then the violin, each for 6 seconds. When a particular instrument plays, the cartoon character holding that instrument on the screen moves, to ensure that the user always knows which instrument is playing. The user's task is to remember the correct order of instruments, and to indicate that order by using the appropriate series of gestures. For instance, after hearing the set of four excerpts in the example above, the user would then mimic gestures for the trumpet, piano, guitar, and violin in order to obtain a perfect score (see the screenshot in Figure~\ref{fig:screenshot}). 
The recall task has been developed to support mental function, because cognitive training, including memory tasks, has been shown to contribute to the maintenance and improvement of cognitive functioning in at-risk elderly people~\cite{ngandu20152}.
Although every melody is comprised of four excerpts in this prototype, more excerpts and instruments will be added in future implementations to increase gameplay difficulty.

\subsection{Customized music for gesture exercises}
Because of the game's specific requirements (e.g., the melody alternating between four instruments, music at the appropriate tempo for the elderly, etc), we composed music for the game rather than using existing music. 
The musical sequences are comprised of four two-bar long phrases, each played by a different instrument, that form a complete 8-bar melody. The order of instruments is randomly varied across musical sequences, and every instrument occurs exactly once in the sequence.

\subsection{Smart gesture detection}

The gesture detection module implements a number of rules that determine which instrument the user is currently mimicking. As described above, the user learns how each instrument should be played during the tutorial phase. 

A ruleset is created for each gesture (trumpet, piano, guitar, or violin playing)\footnote{\dorien{The four rulesets are available at \url{http://katagres.com/musicgesturesgame}}}. These algorithms implement a number of threshold parameters ($a, b, c, d, e, d, f, h$) to make the gesture detection more accurate. The exact settings of these parameters were determined using trial and error. The Kinect coordinates of specific joints of the player are collected through the Kinect sensor as $X_{ij}$, $Y_{ij}$, and $Z_{ij}$, whereby $i \in \{L, R\}$ with $L$ being left and $R$ right, and $j \in \{E, S, D, H\}$ with $E$ for elbow, $S$ for shoulder, $D$ for head and $H$ for hand.

Algorithm~\ref{alg:trumpet}, for instance, describes the rules for determining whether the user correctly mimics the gestures required to play the trumpet, checking whether 1) both hands are above the elbows, 2) both elbows are below the shoulders, 3) the left and right hand are within a close range of each other, 4) the hands are below the middle of the head, and 5) the hands are within a particular range away from the head. When all these checks prove valid, the trumpet playing gesture is detected. 
We chose to focus on these particular four instruments because they require: 1) processing and remembering how each instrument is played, 2) bilateral movements that require coordination (e.g. bringing hands together for playing trumpet), and 3) each instrument gesture is unique and requires a distinct set of movements from the user. 


\begin{algorithm}
 \KwData{Kinect joint coordinates; $a, b$ threshold constants}
 \KwResult{Trumpet playing detected: True or False}
  \eIf{
  ($Y_{LH} > Y_{LE}$) and ($Y_{RH} > Y_{RE})$ and\\
  $(Y_{LS} > Y_{LE})$ and $(Y_{RS} > Y_{RE})$ and \\
  $|(X, Y, Z)_{RH}  (X, Y, Z)_{LH}| < a$ \\
  $|X_{RH}  X_{D}| < b$ and \\
  $|X_{LH}  X_{D}| < b$  
  }{
  True\;  }{
  False\;
  }
 \caption{Trumpet gesture detection module }
 \label{alg:trumpet}
\end{algorithm}

\vspace{-.3cm}
\section{Future research and evaluation}


In the future, we plan to run a randomized controlled trial (RCT) to test the efficacy of our system, similar to the one described in~\cite{agres2017icot}. During an RCT, we will measure both the cognitive abilities and physical activity level of the participants through a dedicated set of assessments. 
We will use a standard clinical assessment tools for measuring cognitive function in elderly, such as the Mini Mental State Exam (MMSE,~\cite{tombaugh1992mini}) or the Montreal Cognitive
Assessment (MoCA,~\cite{nasreddine2005montreal}).
For measuring the \emph{physical activity} of the user, we will employ a tool such as the Physical Activity Scale for the Elderly (PASE)~\cite{washburn1993physical}, and plan to incorporate automatic assessment of
users' range of motion and kinematics using the Kinect sensor. Before the RCT, the game will be enhanced to include more levels, instruments, and advanced gesture detection. The motivational aspects of music may also be enhanced by incorporating music as a reward mechanism, e.g., players will unlock new musical examples or styles as they progress. \looseness=-1 



\section{Conclusion}

We have presented a prototype serious game as preventive medicine for the elderly. 
Our eHealth game aims to 1) leverage the health-related aspects of music (e.g., improve mental health and engage users), 2) increase motivation and compliance with doctor's instructions to remain physically active in old age, 3) encourage the elderly to make a series of coordinated, bilateral upper limb exercises (because coordinated, dissociative movements can benefit both motor ability and cognitive health), and 4) support cognitive function through a real-time interactive memory task. The prototype is part of a larger suite of games for eHealth that includes games for strengthening and rehabilitation~\cite{beveridgerhythmic,agres2017icot}. \looseness=-1




\footnotesize
\bibliographystyle{ACM-Reference-Format}
\bibliography{paper} 

\end{document}